\documentclass[12pt]{article}
\usepackage{natbib, setspace,lscape,longtable}
\usepackage{epsfig, color, verbatim, graphicx}
\usepackage{mathrsfs,amsmath,amsthm,amssymb}

\numberwithin{equation}{section}
\newtheorem{theorem}{Theorem}
\newtheorem{lemma}{Lemma}

\newtheorem{proposition}{Proposition}
\newcommand{\scsection}[1]
    {\begin{center}
        {\bf \large #1}
    \end{center}
}

\newcommand{\mbf}{\mathbf}

\def\beq{\begin{equation}}
\def\eeq{\end{equation}}
\def\beqr{\begin{eqnarray}}
\def\eeqr{\end{eqnarray}}
\def\beqrs{\begin{eqnarray*}}
\def\eeqrs{\end{eqnarray*}}
\def\bet{\begin{theorem}}
\def\eet{\end{theorem}}
\def\bel{\begin{lemma}}
\def\eel{\end{lemma}}
\def\bep{\begin{proposition}}
\def\eep{\end{proposition}}
\def\bg{\begin{figure}[tbph]\begin{center}}
\def\eg{\end{center}\end{figure}}

\def\bc{\begin{center}}
\def\ec{\end{center}}
\def\hat{\widehat}
\def\tilde{\widetilde}
\def\bar{\overline}

\def\1{\mathbf{1}}

\def\mQ{\mathcal Q}
\def\mR{\mathbb{R}}
\def\mS{\mathcal S}
\def\mQ{\mathcal Q}

\def\mE{\mathcal E}

\def\mX{\mathbb{X}}

\def\mY{\mathbb{Y}}
\def\var{\mbox{var}}
\def\cov{\mbox{cov}}
\def\argmax{\mbox{argmax}}
\def\FDP{ \mathrm{FDP} }
\def\FDR{ \mathrm{FDR} }
\def\FAT{ \mathrm{FAT}}
\def\PFA{ \mathrm{PFA}  }

\def\bs{\boldsymbol}
\def\tr{\mathrm{tr}}
\def\u{\mathrm{u}}
\def\o{\mathrm{o}}
\def\ID{\mathrm{I}}
\def \mI {\mathcal{I} }
\def\mbf{\mathbf}

\begin{document}
\begin{center}
{\bf\Large A Factor-Adjusted Multiple Testing Procedure with Application to Mutual Fund Selection}\\

\bigskip
Wei Lan and Lilun Du
%\footnotetext[2]{Wei Lan is from the Statistics School and Center of Statistical Research, Southwestern University of Finance and Economics (E-mail: facelw@gmail.com). Lilun Du is the
%corresponding author and an assistant professor of the Department of ISOM, Hong Kong University of Science and Technology (E-mail: dulilun@ust.hk). Wei Lan's research was supported
%by National Natural Science Foundation of China (NSFC, 11401482, 71532001).
%Lilun Du's research was partially supported by IGN15BM04, SBI16BM01, and Hong Kong RGC ECS26301216. The authors are grateful to the Editor, the AE, and two anonymous referees for their insightful comments and constructive suggestions.}  \\
%
%

{\it Southwestern University of Finance and Economics, and Hong Kong University of Science and Technology}  \\

%This Version: \today
\end{center}

\begin{abstract}
In this article, we propose a factor-adjusted multiple testing (FAT) procedure based on factor-adjusted $p$-values in a linear factor model involving
some observable and unobservable factors, for the purpose of selecting skilled funds in empirical finance \citep{Barras:Scaillet:Wermers:2010}.
The factor-adjusted $p$-values were obtained after extracting the latent common factors by the principal component method \citep{Wang:2012}.
Under some mild conditions, the false discovery proportion can be consistently estimated even if the idiosyncratic errors are allowed to be weakly correlated across units. Furthermore, by appropriately setting a sequence of threshold values approaching zero, the proposed $\FAT$ procedure enjoys model selection consistency.
Extensive simulation studies and a real data analysis for selecting skilled funds in the U.S. financial market are presented to illustrate the practical utility of the proposed method.\\

\noindent {\bf Key Words:} Factor Model; False Discovery Proportion; Fund Performance; High Dimensional Data;
Multiple Testing

\end{abstract}

\newpage
\section{Introduction}\label{Sec-1}

Due to the rapid development of fund of funds (FoF),
how to search for outperforming funds among a large pool of candidates
has attracted much attention from both investors and academic researchers \citep{Barras:Scaillet:Wermers:2010,Fama:French:2010}.
Although the literature claims that active investment should be a negative sum
game after costs and documents negative average fund alphas, recent studies have indicated that some of the funds
indeed display stock-picking ability (see, e.g., Barras et al., 2010), which means that locating funds with significant stock-picking ability is not always a wild goose chase.
In the mutual fund literature,
it is commonly found that funds with significant stock-picking ability
outperform others. Therefore, investors usually
sort the funds according to some particular fund performance measure, such as
the Sharpe ratio or Jensen's $\alpha$, according to their past returns,
and invest more in funds that have recently performed
well with the expectation that they will continue to do so in the future.

Whether the funds with the highest past performance continue to produce higher returns in the future has received much attention in the literature (see, e.g., Carhart, 1997).
To date, there is still no clear evidence that the funds' performance can be persistent. \cite{Fama:French:2010} noted that fund performance persistency tests all have a critical limitation, as they are all based on ranking funds according to their short-term past performance. As a consequence, the allocation of funds into winner funds and loser funds may be largely due to noise or error, which casts doubt on the usefulness of the existing performance persistency tests.
To take into account the possible noise, one possible solution is to
assess the relative effect of luck (noise) and skill (signal) in determining each fund's performance \citep{Cornell:2009}.
Consequently, how to identify those funds whose performance is largely due to skill rather than luck is a topic of practical demand.

Intuitively, the funds' returns are quite hard to reproduce in the
future if they are largely produced due to noise and luck. To this end, it is natural to wonder how many fund managers possess true stock-picking ability.
According to the definition of \cite{Barras:Scaillet:Wermers:2010}, a fund can be regarded as a skilled fund if it has a positive risk-adjusted return net of costs, i.e., the funds that have managers with stock-picking
ability sufficient to cover the trading cost and expenses.
Of course, we cannot observe the true risk-adjusted return of every fund in the population. Therefore, to identify skilled
funds, a simple $t$-test can be applied to each fund to assess whether it has a positive risk-adjusted return. From a statistical point of view, this amounts to
conducting the following $N$ simultaneous hypothesis testing problems:
\[
H_{0i}: \mu_i=0   \quad  \mbox{versus}  \quad  H_{1i}: \mu_i \ne 0,\  i=1, \ldots, N,
\]
where $\mu_i$ is the intercept of fund $i$ specified in the following model:
\[ Y_{i j}=\mu_i+\bs \beta_i^\top \bs{X}_j +\varepsilon_{i j}. \]
Here, $Y_{i j}$ is the excess return of fund $i$ at time $j$, $\bs{X}_j$ are the systematic risk factors related to
some specific linear pricing model, such as the capital asset pricing model (CAPM) of
\cite{Sharpe:1964}, and the three-factor model of \cite{Fama:French:1993}.
Specifically, for the CAPM, $\bs{X}_j$ represents the market index, while for the Fama-French three-factor model, $\bs X_j$ contain three predictors, namely, the market index, the size effect and the value effect, respectively;
more rigourous definitions of the notations are presented in Section~\ref{Sec-2}.

According to the Berk and Green equilibrium \citep{Berk:Green:2004}, only a very small proportion of $\mu_i$s are nonzero.
To identify the sparse positive $\mu_i$ among a large number of simultaneous hypothesis testing problems,
\cite{Barras:Scaillet:Wermers:2010} suggested controlling for
the false discovery rate (FDR) \citep{Benjamini:Hochberg:1995}
instead of familywise type I error, and discussed its usefulness for mutual fund selection. Note that the legality of such a procedure largely depends on one critical assumption, i.e., the test statistics for testing each $H_{0i}: \mu_i=0$ are weakly correlated; otherwise, the FDR cannot
be controlled at the nominal level \citep{Storey:Taylor:Siegmund:2004, Fan:Han:Gu:2012}.
The weak dependence assumption of the test statistics for testing $H_{0i}: \mu_i=0$
is equivalent to assuming that almost all of the variation in the mutual fund returns can be captured by the aforementioned linear asset pricing models, so that
$\varepsilon_{i j}$s are weakly correlated across different funds. However, it is remarkable that whether the aforementioned linear asset pricing models are
adequate to explain fund returns has been a highly controversial topic in the empirical finance over the past 20 years (see, e.g., Kleibergen and Zhan, 2013),
which calls into question the usefulness of the
procedure proposed by \cite{Barras:Scaillet:Wermers:2010}, and some adjustments are inevitable.

To make a possible adjustment, we first need to understand under what circumstances
the random noise in $\varepsilon_{i j}$ is strongly correlated.
Let $\Sigma_{\mE}\in\mR^{N\times N}$
be the covariance matrix of $\mE_j=(\varepsilon_{1j}, \ldots, \varepsilon_{Nj})^\top$.
According to the definition of \cite{Chamberlain:Rothschild:1983}, when the observed systematic risk factors
can fully explain the fund returns asymptotically, we can assume that $\lambda_{\max}(\Sigma_{\mE})<\infty$, with
$\lambda_{\max}(\Sigma_{\mE})$ being the largest eigenvalue of $\Sigma_{\mE}$. In contrast, if the linear pricing model fails to explain the fund returns sufficiently, we can expect that $\lambda_{\max}(\Sigma_{\mE})\rightarrow\infty$.
This is a structure that can be easily explained if the random noise $\varepsilon_{ij}$ admits an approximate factor
structure \citep{Chamberlain:Rothschild:1983}.
In other words, there exist some unobservable systematic risks that are still not
captured by commonly used linear pricing models, so that the funds' excess returns follow a multi-factor error structure involving some observable and unobservable factors. Under such a multi-factor error structure model,
the resulting test statistics share the latent systematic risks, which tend to be highly correlated.
Accordingly, the classical method proposed in \cite{Benjamini:Hochberg:1995} and \cite{Storey:Taylor:Siegmund:2004} is no longer applicable.
To overcome this problem, some recent developments in multiple testing fields have tried to utilize the dependence structure to improve the efficiency.
For instance, \cite{Genovese:Roeder:Wasserman:2006} incorporated the information related to each hypothesis into the procedure of \cite{Benjamini:Hochberg:1995} through weighting the $p$-values, and further showed that the power improved and the FDR was controlled at the nominal level. \cite{Sun:Cai:2009} developed a multiple testing procedure based on a compound decision theoretical framework to exploit the dependence structure and improve the test efficiency. Unfortunately, these procedures are not applicable to our case due to the strong dependence among $\Sigma_{\mE}$. To improve the efficiency and to incorporate the strong dependence information, some appropriate factor-adjusted procedures are needed for the multiple testing.

Motivated by the practical demand, our aim is to modify the method of \cite{Barras:Scaillet:Wermers:2010} via a factor-adjusted multiple testing $(\FAT)$ procedure. Statistically, our interest is in simultaneously testing whether the intercept of each unit is equal to zero in a linear
factor model setting involving some observable and unobservable
factors, while the idiosyncratic errors are weakly correlated.
Under this framework, we first construct a factor-adjusted $p$-value for each test
based on a factor-adjusted test statistic obtained by the principal component method
for extracting the latent common factors. The factor-adjusted $p$-values are then used to implement the multiple testing procedure of
\cite{Storey:Taylor:Siegmund:2004} and \cite{Barras:Scaillet:Wermers:2010}.
Accordingly, we refer to this adjustment as the FAT procedure.
We show theoretically that under some mild conditions, the $\FAT$ procedure can consistently estimate the true false discovery proportion (FDP).
Furthermore, by appropriately setting a sequence of threshold values approaching zero,
the proposed $\FAT$ procedure can achieve consistent model selection
under some sparsity and regularity conditions.  All of the theoretical results are
further confirmed by extensive numerical studies.

The rest of this article is organized as follows.
 The model, notations, and technical conditions are introduced in Section~\ref{Sec-2}.
Theoretical justification of the $\FDR$ estimate and the model selection consistency are discussed in Section~\ref{Sec-3}.
Extensive numerical experiments and a real data analysis are presented in Section~\ref{Sec-4} and Section~\ref{Sec-5}, respectively.
Section~\ref{Sec-6} concludes the article with a short discussion.
All of the technical details are relegated to a supplementary file.
Throughout the article, we denote by $\| \mbf A\|=\{\tr( \mbf A^\top  \mbf A)\}^{1/2}$ the Frobenius norm of any arbitrary matrix $\mbf A$.

\section{Methodology} \label{Sec-2}

\subsection{Notations and assumptions}

We assume that there are $N$ units, and each unit has $T$ observations.
Let $Y_{i j}\in\mR^1$ be the response variable of unit $i$ at time $j$, and let
$\bs{X}_j=(X_{1 j},\ldots, X_{p j})^\top\in\mR^p$ for $p\geq 0$ be the observed explanatory variables.
Here, $p=0$ is associated with the case where no explanatory variables are involved.
Throughout the article, unless explicitly stated otherwise, we assume that
$N\gg T$ and $T$ tends to infinity for asymptotic behavior, while $p$ is fixed.
We next consider the following multivariate linear regression,
\beq  \label{2.1}
Y_{i j}=\mu_i+\bs\beta_i^\top  \bs{X}_j+\varepsilon_{ij},
\eeq
where
$\bs \beta_i=(\beta_{i1},\ldots, \beta_{ip})^\top\in\mR^p$, $\varepsilon_{ij}$ is the corresponding random error, and $\mu_i$ is the intercept.
Define $\mY_i=(Y_{i1}, \ldots, Y_{iT})^\top\in\mR^T$ and let $\mY=(\mY_1, \ldots, \mY_N)^\top \in\mR^{N \times T}$ be the response matrix. Let $\mX=(\bs{X}_1, \ldots, \bs{X}_T)^\top \in \mR^{T\times p}$ be the design matrix, and
$\bs\mu=(\mu_1, \ldots, \mu_N)^\top\in\mR^N$ collects all of the intercepts for each unit. In addition, we assume that the random errors $\mE_j=(\varepsilon_{1 j}, \ldots, \varepsilon_{Nj})^\top\in\mR^N$ are
independent and identically distributed with a mean of $\boldsymbol{0}$ and a covariance matrix $\Sigma_{\mE }$ in the time domain, and construct the error matrix as $\mE=(\mE_1, \ldots, \mE_T)\in\mR^{N \times T}$.
Without loss of generality, we assume that all of the explanatory variables have been
appropriately centralized such that $E(X_{lj})=0$ for every $1\le l \le p, 1\le j \le T$. To further model the dependence structure of $\Sigma_{\mE}$,
we assume that $\mE_j$ admits the following latent factor structure
\citep{Fan:Liao:Mincheva:2013},
\beq  \label{2.2}
\mE_j=\bs \Gamma \bs{Z}_j+\bs\eta_{.j},  \quad  j=1, \ldots, T,
\eeq
where $\bs{Z}_j=(Z_{1j},\ldots, Z_{rj})^\top\in\mR^r$ is the low dimension of $r$ unknown common factors
with the identification restriction $\cov( \bs{Z}_j)=\mbf{I}_r$;
$\bs \Gamma=(\bs\gamma_1, \ldots, \bs\gamma_N)^\top   \in   \mR^{N\times r}$ are the
unknown factor loadings; and $\bs\eta_{.j}=(\eta_{1j},\ldots, \eta_{Nj})^\top\in\mR^N$ is the random noise that is independent of $\bs Z_j$ and $\bs X_j$. For simplicity, we assume there are no interaction effects between the explanatory variables $\bs X_j$ and the latent factors $\bs Z_j$.
To further model the test-specific variations,
we assume that $\bs\eta_{.j}$ is normally distributed and weakly dependent such that, for some positive constant $\delta>0$,
\begin{eqnarray}
&& \bs \eta_{.j}  \sim^{d}  \mathbb{N} (\boldsymbol{0}, \Sigma_{\bs \eta}) \quad \mbox{with}\ \Sigma_{\bs \eta}=(\sigma_{\bs \eta, i_1i_2}), \cr
&&N^{-2+\delta}\sum_{i_1\ne i_2} \big| \rho_{\bs \eta, i_1i_2}\big|=O(1),  \label{2.3}
\end{eqnarray}
where $\mathbb{N}(\cdot,\cdot)$ stands for multivariate normal distribution,
and $\rho_{\bs \eta, i_1i_2}=\sigma_{\bs \eta, i_1i_2}/(\sigma_{\bs \eta, i_1i_1} \sigma_{\bs \eta, i_2i_2})^{1/2}$ is the correlation coefficient between $\eta_{i_1 j}$ and $\eta_{i_2 j}$. It is
worthy noting that \cite{Fan:Han:Gu:2012} introduced a similar condition to that in \eqref{2.3} (replacing the pairwise correlation with its covariance component) for weakly dependent normal random variables, to explore the extent to which their approximated FDP could be consistently estimated. These two definitions are essentially equivalent when the eigenvalues of $\Sigma_{\bs \eta}$ are bounded away from zero and infinity.

Under the multi-factor error structure of (\ref{2.2}), model \eqref{2.1} can be further written as
\beq Y_{i j}=\mu_i+\bs \beta_i^\top \bs X_j+\bs \gamma_i^\top \bs Z_j+\eta_{ij}. \label{2.4}  \eeq
The main focus of the current article is to simultaneously test the following hypotheses:
\begin{equation} \label{2.5}
H_{0i}: \mu_i=0   \quad  \mbox{versus}  \quad  H_{1i}: \mu_i \ne 0,\  i=1, \ldots, N,
\end{equation}
under the model setup \eqref{2.4} based on the observations $\mY$ and $\mX$,
which are addressed in the next subsection.

\subsection{Factor-adjusted multiple testing procedure}

We first consider the testing problems in \eqref{2.5} by
ignoring the dependence structure of $\mE_j$. Using the traditional ordinary
least squares (OLS) estimation method based on the
 observed data $\{\mY, \mX\}$, we can obtain the estimated intercept
 $\hat{\bs \mu}^\top=(\1^\top \mQ\1)^{-1} \1^\top \mQ(\mX) \mY^\top$,
where the projection operator $\mQ(\mX)=\mbf I_T-\mX(\mX^\top\mX)^{-1}\mX^\top$ and $\1=(1,\ldots, 1)^\top\in\mR^T$ is a vector of $1$s of dimension $T$.
The asymptotic distribution of $\hat{\bs \mu}$ is
\begin{equation*}
(\1^\top \mQ \1)^{1/2} \hat{\bs \mu} \rightarrow^d \mathbb{N}\big( (\1^\top \mQ \1)^{1/2}\bs\mu, \Sigma_{\mE} \big), \ \mbox{as} \
T, N \rightarrow \infty.
\end{equation*}
Accordingly, the unadjusted test statistic
for the $i$-th hypothesis is defined as $T_i^{\u}={\hat{\mu}_i }/{ s.e.(\hat{\mu}_i )}$
with $s.e.(\hat{\mu}_i )=(\1^\top \mQ(\mX) \1)^{-1/2} (\hat{\sigma}_{\mE, ii} )^{1/2}$, where $\hat{\sigma}_{\mE, ii}$ is the $i$-th diagonal element of $\hat{\Sigma}_{\mE}=T^{-1}\mQ(\tilde{\mX})\mY^\top\mY \mQ(\tilde{\mX})$ with $\tilde\mX=(\1, \mX)$. By normal approximation, the unadjusted $p$-values are formulated as $P_i^{\u}=2 \Phi(-|T_{i}^{\u} |)$, for $i=1,\ldots,N$.
In summary, the unadjusted test statistics follow a multivariate normal distribution with a strongly correlated covariance structure, which can be perfectly cast into the principle factor approximation (PFA) estimation framework developed by \cite{Fan:Han:Gu:2012}. Although their method is accurate in terms of consistently estimating the FDP, the rank of the rejected hypotheses remains unchanged, and thus is still inefficient even if they exploit the dependence structure of $\mE$. In this article, we seek to design a new testing procedure to improve the efficiency under the multi-factor error structure in \eqref{2.4}. Toward this end, we propose an alternative $p$-value based on factor-adjusted test statistics, which significantly changes the signal-to-noise ratio for testing the intercepts in \eqref{2.5} and thus the structure of the $p$-values. More importantly, removing the latent common factors from the factor-adjusted $p$-values means that the weak dependence assumption among the factor-adjusted $p$-values should be satisfied, and hence the multiple testing procedure in \cite{Storey:Taylor:Siegmund:2004} can be implemented in the downstream analysis. We elaborate this idea in details as follows.

According to \eqref{2.2}, we can obtain
\begin{equation}  \label{2.6}
\hat{\bs\mu}^\top-\bs\mu^\top=(\1^\top \mQ(\mX)\1)^{-1}\1^\top \mQ(\mX)\mbf{Z}\bs\Gamma^\top +(\1^\top\mQ(\mX)\1)^{-1} \1^\top \mQ(\mX) \bs \eta^\top,
\end{equation}
where $\mbf{Z}=(\bs{Z}_1, \ldots, \bs{Z}_T)^\top\in\mR^{T\times r}$, $\bs \eta=(\bs \eta_{.1}, \ldots, \bs \eta_{.T} ) \in\mR^{N \times T}$. We note that $(\1^\top \mQ(\mX)\1)^{-1} \1^\top \mQ(\mX) \bs \eta^\top$ follows a multivariate normal distribution:
\begin{equation}  \label{2.7}
(\1^\top \mQ(\mX)\1)^{-1/2} \1^\top \mQ(\mX) \bs \eta^\top \sim^d \mathbb{N} \big(0, \Sigma_{\bs \eta}  \big).
\end{equation}
The weak dependence assumption in \eqref{2.3} tells us that after removing the latent factor $\mbf Z$ from $\hat{\bs \mu}$,
the behavior of the resulting test statistics will be analogous to independence. Consequently, we define an Oracle factor-adjusted
test statistic as
\begin{eqnarray}
T_i^o&=&\frac{ (\1^\top \mQ(\mX)\1)^{1/2}\hat{ \mu}_i-(\1^\top \mQ(\mX)\1)^{-1/2} \1^\top \mQ(\mX)  \mbf{Z}\bs\gamma_i   }{ (\sigma_{\bs \eta, ii})^{1/2} } \cr
&=&\frac{(\1^\top \mQ(\mX)\1)^{1/2} \mu_i+(\1^\top \mQ(\mX)\1)^{-1/2} \1^\top\mQ(\mX) \bs \eta_{i.} }{ (\sigma_{\bs \eta, ii} )^{1/2} }.
\qquad  \label{2.8}
\end{eqnarray}
From \eqref{2.7} and \eqref{2.8}, $(T_1^o, \ldots, T_N^o)^\top$ follows a multivariate normal distribution with a mean of $(\1^\top \mQ \1)^{1/2}D(\bs \eta)^{-1/2} \bs \mu$ and a covariance matrix scaled from $D(\bs \eta)^{-1/2}\Sigma_{\bs \eta}D(\bs \eta)^{-1/2}$, with $D(\bs \eta)=\mbox{diag}\{\sigma_{\bs \eta, 11}, \ldots, \sigma_{\bs \eta, NN} \}$. An Oracle factor-adjusted $p$-value for \eqref{2.5} is then defined as  $P^{\o}_i=2\Phi(-|T_i^o|)$.
The plug-in method naturally leads to a factor-adjusted test statistic and its corresponding $p$-value defined as
\begin{eqnarray}  \label{2.9}
\hat{T}_i = \frac{(\1^\top \mQ(\mX)  \1)^{1/2} \hat\mu_i -(\1^\top \mQ(\mX)\1)^{-1/2} \1^\top \mQ(\mX)  \hat{\mbf{Z}}  \hat {\bs \gamma}_i  }
{  (\hat{\sigma}_{\bs \eta, ii})^{1/2} } \quad \mbox{and} \quad
P_i = 2\Phi(-|\hat{T}_i |),
\end{eqnarray}
where $\hat{\sigma}_{\bs \eta, ii}$, $\hat{\mbf{Z}}$, and  $\hat{\bs \gamma}_i$ are some
estimators of the idiosyncratic error variance, unobservable factors, and factor loadings, respectively. The details of $\hat{\sigma}_{\bs \eta, ii}$, $\hat{\mbf Z}$, and $\hat{\bs \gamma}_i$ are discussed in Section~\ref{Sec-3.1}.
%Intuitively, the null distribution of $\widehat{T}_i$ can be approximated by a standard normal distribution, from which the factor-adjusted $p$-value for the $i$-th hypothesis of \eqref{2.5} is then defined as $P_i=2\Phi(-| \widehat{T}_i |)$.

Before we describe our factor-adjusted multiple testing procedure, we introduce some commonly used notations. The sets of indices corresponding to the true null and non-null in \eqref{2.5} are denoted by $\mI_0$, $\mI_1$, respectively.
Let $N_0$ and $N_1$ be the cardinality of $\mI_0$ and $\mI_1$, and define by $\pi_0=\lim_{N \rightarrow \infty}N_0/N$ the asymptotic proportion of the true null. Similar to \cite{Storey:Taylor:Siegmund:2004}, we define the following empirical processes:
\begin{eqnarray}  \label{2.10}
V(t) &=& \# \{ i\in \mI_0: P_i \le t\},  \cr
S(t) &=& \# \{ i\in \mI_1: P_i \le t\},  \quad   \mbox{and} \cr
R(t) &=& \# \{ i\in \{1, \ldots, N\}: P_i \le t \}
\end{eqnarray}
for any $t \in [0,1]$. Then, $V(t), S(t)$, and $R(t)$ are the number of falsely rejected hypotheses, the number of correctly rejected
hypotheses, and the total number of rejected hypotheses, respectively.  The FDP with respect to the threshold $t$
is defined as $\mathrm{FDP}(t)=V(t)/\{R(t)\vee 1\}$ with $R(t)\vee 1=\max\{R(t), 1\}$. The FDR is defined as the expectation of $\mathrm{FDP}$, i.e., $\mathrm{FDR}(t)=E\big\{\mathrm{FDP}(t)\big\}$.
%It is worth mentioning that $V(t)$ is unobserved but realized through an experiment, while $R(t)$ can be observed.
Analogously, $V^{\o}(t)$, $S^{\o}(t)$, $R^{\o}(t)$, $\FDP^{\o}(t)$, and $\FDR^{\o}(t)$ are defined in a similar way as those in \eqref{2.10} by using the Oracle factor-adjusted $p$-values.

As expected, $\hat{T}_i$ will resemble $T_i^o$ to a large extent as long as the latent factors and factor loadings can be estimated with certain accuracy. This, together with the weak dependence structure among $T_i^o$s, motivates us to test \eqref{2.5} using the procedure of \cite{Storey:Taylor:Siegmund:2004}, which is less conservative than the method of \cite{Benjamini:Hochberg:1995}. Specifically, for a pre-chosen level $\alpha$ and a tuning parameter of $\lambda \in [0,1)$,
a data-driven threshold for the $p$-values is determined by
\begin{eqnarray}  \label{2.11}
 t_{\alpha} \big(  \widehat{ \mathrm{FDR} }_{\lambda} \big)= \sup \big\{  0 \le t \le 1:
 \ \widehat{ \mathrm{FDR} }_{\lambda}(t) \le \alpha \big\},
\end{eqnarray}
where $\widehat{ \mathrm{FDR} }_{\lambda}(t)$ is a point estimate of $\mathrm{FDR}(t)$, which is given by
\begin{eqnarray}  \label{2.12}
 \widehat{ \mathrm{FDR} }_{\lambda}(t)=
 \frac{N \hat{\pi}_0(\lambda) t }{R(t) \vee 1}
=\frac{ \hat{\pi}_0(\lambda) t  }{\{R(t) \vee 1\}/N},
\end{eqnarray}
where $\hat{\pi}_0(\lambda)=\big\{N(1-\lambda)\big\}^{-1}\big\{N-R(\lambda)\big\}$ is an estimate of $\pi_0$. A good choice of $\lambda$ should ensure that the alternative hypotheses with the factor-adjusted $p$-values larger than $\lambda$ are negligible.
We then reject the null hypothesis if its $p$-value ($P_i$) is less than or
equal to $t_{\alpha} (  \widehat{ \mathrm{FDR} }_{\lambda} )$.
Hereafter, we refer to \eqref{2.12} as the \textit{estimation approach} for $\mathrm{FDR}$ and \eqref{2.11} as the \textit{controlling approach} for $\mathrm{FDR}$. The data-driven threshold \eqref{2.11} together with the point estimate method \eqref{2.12} for the FDR comprises the FAT procedure.

\subsection{Connections with and differences from existing methods} \label{Sec-3.3}

The idea of adjusting the dependence effect using a factor model under a multiple testing framework is not new; it has been tentatively studied by \cite{Leek:Storey:2008}, \cite{Friguet:Kloareg:Causeur:2009}, and \cite{Fan:Han:Gu:2012}.
The proposed FAT procedure differs from these previously described methods in the following two respects.

\begin{center}
{\it  (1.) Relation to \cite{Leek:Storey:2008} and \cite{Friguet:Kloareg:Causeur:2009}}
\end{center}
The main aim of \cite{Leek:Storey:2008} and \cite{Friguet:Kloareg:Causeur:2009}
is to simultaneously test the significance of
the regression coefficients $\bs \beta_i$ for $i=1, \ldots, N$, while our aim is to
simultaneously test the significance of the intercepts motivated by mutual fund selection in empirical finance.  By definition, in \cite{Leek:Storey:2008}, the linear space spanned by the latent random vectors that captures the dependence among the tests is termed the \textit{dependence kernel}, which has two specifically scientific applications, i.e.,
the spatial dependence typically assumed in brain-imaging data and the latent structure due to relevant factors not included in
biological studies. In contrast, our latent factor structure $\bf Z$ has implications for unobserved systematic risks, and is thus relevant to empirical finance. More importantly, the unique feature for testing high dimensional intercepts alleviates
the \textit{confounding} phenomenon between the observed and unobserved systematic risks, whereas the existing procedures for
testing regression coefficients associated with observed explanatory variables are prone to generating spurious signals due to interactions between the observed and unobserved common factors.
In addition, \cite{Leek:Storey:2008} and \cite{Friguet:Kloareg:Causeur:2009} used some variants of EM-type algorithms to estimate the number of latent factors, the factor loadings, and the latent factors.
Such a method is demonstrated to be quite
useful in simultaneous point estimation. Nevertheless, the resulting estimators
do not have explicit solutions, which poses more challenges when investigating the effect of estimation errors
on the subsequent testing procedure. As an alternative, we propose to estimate the factor number, the latent factors, and the factor loadings using the principal component method \citep{Wang:2012}; see Section~\ref{Sec-3.1} for details.
We show theoretically that such a simple procedure can consistently estimate the true FDP even when
the idiosyncratic errors are allowed to be weakly correlated. Consequently, our procedure
has formal theoretical justifications under much weaker conditions,
compared with the methods of
\cite{Leek:Storey:2008} and \cite{Friguet:Kloareg:Causeur:2009}.

\begin{center}
{\it  (2.) Relation to \cite{Fan:Han:Gu:2012} and \cite{Fan:Han:2013}}
\end{center}
To deal with an arbitrary dependence between test statistics,
\cite{Fan:Han:Gu:2012} and \cite{Fan:Han:2013} proposed a novel method based on principal factor approximation, which successfully subtracts the common dependence and significantly weakens the correlation structure. By applying eigenvalue decomposition to the covariance matrix of the test statistics, the test statistics can be represented as a factor model; this is subtly different from
the factor structure of the raw data considered in this article, and from the methods of \cite{Leek:Storey:2008} and \cite{Friguet:Kloareg:Causeur:2009}.
The principal factor approximation is demonstrated to be very appealing for the purpose of estimating the FDP.
Nevertheless, the meaning of the latent factors captured by test statistics is quite hard to interpret.
In contrast, in our model, the latent factors can be regarded as some unobserved systematic
risks in the financial market and thus have apparent economic meaning.
Moreover, it is useful that \cite{Fan:Han:Gu:2012}
designed a simple method to select the number
of latent factors. However,
%as illustrated by our simulation studies in Section~\ref{Sec-4},
the selection of the number of latent factors is volatile with the choice of threshold values and the strength of signals.
To fix this issue, we apply an eigenvalue ratio
criterion in the spirit of \cite{Wang:2012} to select the number of
latent factors, and show theoretically that the criterion can
select the number of latent factors consistently when the unknown intercepts are quite sparse. For comparison, we also present the methods of \cite{Fan:Han:Gu:2012} and \cite{Fan:Han:2013}
in the simulation studies in Section~\ref{Sec-4}.

\section{Theoretical Framework}\label{Sec-3}

In this section, we first theoretically justify that the factor-adjusted $p$-values under the true null
satisfy a similar weak dependence assumption to that of \cite{Storey:Taylor:Siegmund:2004},
under which the asymptotic properties of the \textit{estimation approach} for the FDR in \eqref{2.12}
can be established in a sparse setting with $\pi_0 =1$ for any given threshold $t$.
Subsequently, when $t \rightarrow 0$, we turn to derive a sequence of threshold values $t_T \rightarrow 0$
such that $V(t_T)  \rightarrow 0$ and $S(t_{T} )/N_1 \rightarrow 1$. Accordingly, the $\FAT$ procedure enjoys
consistent model selection.

\subsection{Weak dependence of the factor-adjusted $p$-values under the true null} \label{Sec-3.1}

As discussed in \cite{Storey:Taylor:Siegmund:2004}, the assumption of weak dependence among the $p$-values plays
an essential role in controlling for the FDR. To verify that our factor-adjusted $p$-values share similar properties,
we propose a two-stage approach to justify the weak dependence of the null factor-adjusted $p$-values. In the first stage, we supply a sufficient condition on the estimators of the latent factors and the variance of the idiosyncratic error,
under which the empirical distribution of the null factor-adjusted $p$-values will convergence to the uniform distribution almost surely.
%This type of approximation is similar to the approximate expression of $\FDP$ derived in \cite{Fan:Han:Gu:2012}.
In the second stage, we estimate the unknown factors and factor loadings
using the principal component method, and further show that these estimators satisfy the sufficient condition in Stage I. Combining these two stages, the weak dependence assumption of the null factor-adjusted $p$-values can be justified and the approximated distribution function (i.e., the uniform distribution) can be directly utilized in the downstream multiple testing.

\begin{center}
{ \it  Stage I: A sufficient condition for the weak dependence of the factor-adjusted $p$-values under the true null}
\end{center}

To assess the asymptotic property of $V(t)$, we evaluate it
based on the Oracle factor-adjusted test statistic $\widetilde T_i$ and the given estimators
$\hat{\sigma}_{\bs \eta, ii}$, $\hat{ \mbf Z}$, and $\hat{\bs \gamma}$ as follows:
%\begin{small}
\begin{align}
V(t)=&\sum_{i \in \mI_0} \ID(P_i \le t)=\sum_{i\in \mI_0 } \ID(|\hat T_i|>-z_{t/2}) \cr
=&\sum_{i\in\mI_0 } \Big\{ \ID(\hat T_i>-z_{t/2})+\ID(\hat T_i<z_{t/2})\Big\}   \cr
=&\sum_{i\in\mI_0 } \Bigg[ \ID\Big\{T_i^o> -z_{t/2} (\hat \sigma_{\bs \eta, ii})^{1/2} /(\sigma_{ii, \bs \eta})^{1/2}+Bias \Big\} \cr
& \quad  \quad +\ID\Big\{T_i^o <z_{t/2} (\hat \sigma_{\bs \eta, ii})^{1/2} /(\sigma_{\bs \eta, ii})^{1/2}+Bias\Big\}\Bigg], \label{3.1}
\end{align}
%\end{small}
where $Bias=(\1^\top\mQ(\mX)\1)^{-1/2}\1^\top\mQ(\mX)(\hat{\mbf Z} \hat{\bs \gamma}_i- \mbf{Z} \bs \gamma_i)/(\sigma_{\bs \eta, ii})^{1/2}$ and $z_{t/2}=\Phi^{-1}(t/2)$ is the $t/2$ lower quantile of a standard normal distribution.
With the normality properties from $\{T_i^o, i=1, \ldots, m\}$ and the weak dependence among them, $V(t)$ can be further approximated by the following proposition.
\begin{proposition}  \label{Proposition-1}
Under the assumption \eqref{2.3}, for any
estimators $\hat{\bs \gamma}_i$, $\hat{\mbf Z}$, and $\hat\sigma_{\bs \eta, ii}$ that
satisfy $\max_{i\in\mI_0} |\hat\sigma_{\bs \eta, ii}-\sigma_{\bs \eta, ii}|\rightarrow_p 0$ and $\max_{i\in\mI_0}|(\1^\top\mQ(\mX)\1)^{-1/2}\1^\top\mQ(\mX)(\hat{\mbf Z} \hat{\bs \gamma}_i-\mbf{Z}\bs \gamma_i)|\rightarrow_p 0$, we can obtain that, with Probability tending to one,
\begin{small}
\begin{eqnarray}  \label{3.2}
&&\lim_{N \rightarrow \infty } N_0^{-1} V(t)- t=0.
\end{eqnarray}
\end{small}
\end{proposition}
\noindent
The result of Proposition~\ref{Proposition-1} indicates that
the factor-adjusted $p$-values under the true null satisfy the weak dependence assumption of \cite{Storey:Taylor:Siegmund:2004} if the estimators $\hat\sigma_{\bs \eta, ii}$, $\hat{\mbf Z}$, and $\bs{\hat\gamma_i}$ satisfy $\max_{i\in \mI_0} |\hat{\sigma}_{\bs \eta, ii}-\sigma_{\bs \eta, ii}| \rightarrow_p 0$ and $\max_{i \in \mI_0 }|(\1^\top\mQ(\mX)\1)^{-1/2}\1^\top\mQ(\mX)(\hat{\mbf Z}\hat{\bs \gamma}_i-\mbf Z\bs\gamma_i) |\rightarrow_p 0$,
which is essential for deriving the consistency of the estimated FDP.
To make our procedure operational, we estimate the latent factors and the variance of the idiosyncratic error using the principal component method \citep{Wang:2012}. Fortunately, these estimators indeed satisfy the condition assumed in Proposition~\ref{Proposition-1}.

\begin{center}
{\it  Stage $\mathrm{II}$: Principal component analysis and its unform consistency}
\end{center}

We estimate $\mbf Z$ and $\bs \Gamma$ using the principal component method
\citep{Wang:2012}. Specifically,
we first extract the effect of the observed explanatory variables $\mX$ by regressing $\mY_i$ on $\mX$, and obtain the residual as
$\hat\mE^\top=\mQ(\mX)\mY^\top=\mQ(\mX)\1\bs \mu^\top+\mQ(\mX)\mE^\top$.
We next define $\hat\lambda_e$ to be the $e$-th largest eigenvalue of
$(TN)^{-1}\hat \mE^\top \hat \mE$, and $\hat{\bs\varrho}_e$ to be the corresponding eigenvector. Consequently, we set $\hat{\mbf Z}= T^{1/2}(\hat{\bs \varrho}_1, \ldots, \hat{\bs \varrho}_{\hat r})$, and $\bs \Gamma^\top$ can be estimated by
$(\hat{\mbf Z}^\top \hat {\mbf Z})^{-1}\hat{\mbf Z}^\top \hat\mE ^\top$. Based on the estimators $\hat{\bs \Gamma}$ and $\hat{\mbf Z}$,
we can obtain the estimated random error as $\hat{\bs \eta}^\top=\mQ(\hat{\mbf Z}) \hat\mE^\top$. Subsequently, $\sigma_{\bs \eta, ii}$ can be estimated by
$\hat{\sigma}_{\bs \eta, ii}=T^{-1}\hat{\bs \eta}_{i.}^\top \hat{\bs \eta}_{i.}$.
Moreover, we define notations $\tilde\lambda_e$, $\tilde{\bs \varrho}_e$, $\tilde{\bs \Gamma}$, $\tilde{\mbf Z}$, and $\tilde\sigma_{\bs \eta, ii}$
as the associated estimators based on the extracted error $\mQ(\mX)\mE^\top$, while
$\bar\lambda_e$, $\bar{\bs \varrho}_e$, $\bar{\bs \Gamma}$, $\bar{\mbf Z}$, and $\bar\sigma_{\bs \eta, ii}$ are the estimators based on the true error $\mE^\top$.
Practically, in the spirit of \cite{Wang:2012},
$\hat r$ can be selected by maximizing the
eigenvalue ratios as $\hat r=\argmax_{e \leq \pi_{\max}} \hat\lambda_e/\hat\lambda_{e+1}$
with some pre-specified
maximum possible order $\pi_{\max}$. The difference is that some unknown
sparse intercepts $\bs \mu$ are involved in our estimation procedure, which poses more challenges
for investigating the consistency of $\hat r$ by carefully taking into account
the effect of the intercepts. We remark here that the principal orthogonal complement thresholding method in \cite{Fan:Liao:Mincheva:2013} can serve as an alternative to estimate the latent factors and factor loadings, but selecting the ``threshold" involves a heavier computational burden.

To investigate the theoretical property of the above estimators, we first present the following conditions:
\begin{itemize}
\item [(C1)] $\log N\leq C_{\hbar} T^{\hbar}$ for some positive constants $\hbar<1/2$ and $C_{\hbar}>0$.
\item [(C2)]{\it Sparsity}, for any finite positive constant $C_{\bs \mu}$,
\[ \| \bs \mu \|\leq C_{\bs \mu}\min\Big\{N^{1/2}/T^{1/2}, N/(TN_1^{1/2})\Big\}.\]
\item [(C3)] $\sigma_0^2 \le \min_{i} \sigma_{\bs \eta, ii} \le \max_{i}\sigma_{\bs \eta, ii} \le \sigma_1^2$ for some  positive constants $\sigma_0^2$ and $\sigma_1^2$.
\item [(C4)] There exists some positive definite matrix $\Sigma_{\bs \Gamma}$ of dimension $r$ such
that $N^{-1}\bs \Gamma^\top \bs \Gamma \rightarrow \Sigma_{\bs \Gamma}$, with the eigenvalues of $\Sigma_{\bs \Gamma}$ bounded from zero to infinity. In addition, there exists some positive constant $\gamma_{\max}$ such that $\max_i \| \bs \gamma_i \|^2\leq \gamma_{\max}$.
\item [(C5)] $\max_{e \leq r} \|\bar{\bs \psi}_e-\bs\psi_e \|=O_p(N^{-\nu})$ for some positive constant $\nu>0$,
where $\bar{\bs \psi}_i$ and $\bs \psi_e$ are the $e$-th eigenvector of $T^{-1}\mE\mE^\top$ and $\Sigma_{\mE}$, respectively.
%
%\item[(C6)] Assume that $R(t)/N \ge H$ for some $H>0$, as $N\rightarrow \infty$.
\end{itemize}

\noindent
By condition $\mathrm{(C1)}$, the number of units $N$ could increase exponentially
with the sample size $T$, so that
$N$ can be much larger than $T$.
Condition $\mathrm{(C2)}$ can be satisfied if $N_1$ is finite, which is reasonable for mutual fund selection
\citep{Barras:Scaillet:Wermers:2010}.
Condition $\mathrm{(C3)}$ can be satisfied
if the eigenvalues of $\Sigma_{\bs \eta}$ are bounded from zero. Similar
conditions are commonly assumed in the literature; see, for example,
\cite{Wang:2009}. Condition $\mathrm{(C4)}$ is also
widely used in the literature, and can be satisfied if the factor loadings are stationary in the sense that the signals of the latent
factors are comparable and
not weak \citep{Ahn:Horenstein:2013}.
Lastly,  condition $\mathrm{(C5)}$ is directly borrowed from \cite{Fan:Han:2013}, and can be
valid for various structures of $\Sigma_{\mE}$ and large $N$;
for a more detailed discussion of this condition, we refer readers to \cite{Fan:Han:2013}.
Under the above conditions, we demonstrate the following result.

\begin{proposition} \label{Proposition-2}
Suppose the conditions $\mathrm{(C1)}$--$\mathrm{(C5)}$ hold. Then,
\begin{equation}  \label{3.3}
(i)~~\mathrm{P}(\hat r=r)\rightarrow 1.
\end{equation}
If we further assume $\tilde\lambda_{e-1}-\tilde\lambda_e \geq d_N$ for some positive constant $d_N$ and for any
$e=2, \ldots, r$, then we have
\begin{eqnarray}
(ii)~~
&&\max_{i \in \mI_0}|\hat{\sigma}_{\bs \eta, ii}-\sigma_{\bs \eta, ii}|= O_p(\sqrt{\log N/T}), \cr
&&\max_{i \in \mI_0}|(\1^\top\mQ(\mX)\1)^{-1/2}\1^\top\mQ(\mX)(\hat{\mbf Z} \hat{ \bs\gamma}_i-\mbf Z \bs \gamma_i)| \cr
&=&O_p\{ (\log N)^{1/4}/T^{1/2}\}+O_p(N^{-\nu}) \cr
 && +O_p\Big(N^{-1}T\|\bs \mu\|^2\Big)+O_p\Big(N_1^{1/2}N^{-1}T\|\bs \mu\|\Big). \label{3.4}
\end{eqnarray}
\end{proposition}

\noindent
%The proof is provided in Appendix C of a supplementary file.
According to the theoretical results of
Proposition~\ref{Proposition-2}(i), $\hat r$ can be equal to $r$ with a
probability approaching one. Hence, to simplify the technical derivations, we assume that the number of true latent factors $r$ is known. The result of Proposition~\ref{Proposition-2}(ii) indicates that the estimators of the latent factors and the variance of the idiosyncratic error share the property of uniform consistency, thus the sufficient condition assumed on Proposition~\ref{Proposition-1} indeed holds, supporting that the null factor-adjusted $p$-values satisfy the weak dependence assumption when the latent factors are estimated by the principal component method.

\noindent\textbf{Remark 1:}
Note that the initial estimator of the intercepts
$\hat{\bs \mu}$ defined in (\ref{2.6}) is obtained by extracting the effect of
$\mX$. As a result, to further control the effect of latent common factors, we only need to focus on the error after extracting the effect of the observed explanatory variables $\mX$, and it is not necessary to control for the intercepts $\bs \mu$. In fact, if the errors are obtained by controlling for the effects of both $\mX$ and $\bs \mu$, that is, the error is defined as $\mQ(\tilde{\mX})\mY^\top=\mQ(\tilde{\mX})\mE^\top$, the resulting $\FAT$ procedure can lead to incorrect FDR control. There are two reasons which can be simply summarized as follows.
Technically, from \eqref{3.4}, if $\hat{\mbf Z}$ is estimated via the full projection matrix $\mQ(\tilde{\mX})$, $\hat{\mbf Z}\hat{\gamma}_i$ will converge to $\mQ(\tilde{\mX})\mbf Z\bs{\gamma}_i$. This implies that the adjusted term $\1^\top\mQ(\mX)\hat{\mbf Z} \hat{\bs{\gamma}}_i$ in \eqref{2.9} will be approximated as $\1^\top \mQ(\mX)\mQ(\tilde{\mX})\mbf{Z}\bs{\gamma}_i$, a quantity that must be exactly zero by the fact that $\mQ(\mX)\mQ(\tilde{\mX})=\mQ(\tilde{\mX})$ and $\1 \perp \mQ(\tilde{\mX})$.
As a result, the effect of adjustment for the common factors on the test statistics disappears. Second, controlling for the effects of both $\mX$ and $\bs \mu$ involves estimating the intercepts $\bs \mu$ through OLS. When $N$ is much larger than $T$, the resulting estimator of $\bs \mu$ is inconsistent, and this type of
inconsistency can not be adjusted through the projection matrix $\mQ(\mX)$ according to \eqref{2.6}. Because we ignore the information in $\bs \mu$ when estimating the latent factors and factor loadings, the signals in $\bs{\mu}$ will be treated as ``noise" and shifted into the estimation error of $\hat{ \mbf{Z} }$. Consequently, two additional terms (i.e., $N^{-1}T \|\bs{\mu}\|^2$ and $N^{-1}T N_1^{1/2} \|\bs{\mu}\|$) appear in the convergence rate in \eqref{3.4} and in Theorem~\ref{Theorem-1} as shown below. To guarantee the consistency of the estimated FDR,  the signals of $\bs \mu$ are required to be quite weak, i.e., the intercepts are extremely sparse; see also condition $\mathrm{(C2)}$ for details. To remove the sparsity condition, the principal component method can be directly applied to a small set of mutual funds with luck (i.e., $\mu_i=0$) to estimate the latent factors. A similar idea was proposed by \cite{Gagnon-Bartsch:Speed:2012}, but this is beyond the scope of this article and could be considered in future research.

\subsection{$\FDP$ property and model selection consistency}

In this subsection, we discuss the asymptotic property of our FDP estimator when the intercepts are so sparse that condition (C2) holds. Under this setup, the consistency of $\widehat{\FDR}_{\lambda}(t)$ can be readily obtained based on Propositions~\ref{Proposition-1} and ~\ref{Proposition-2}.
\begin{theorem} \label{Theorem-1}
Suppose the assumption~\eqref{2.3} and conditions $\mathrm{(C1)}$--$\mathrm{(C5)}$ hold.
Then, for any $\lambda \in [0, 1)$ and $t>0$, we have
\begin{eqnarray}  \label{3.5}
&& \big| \widehat{\FDR}_{\lambda}(t)-\FDP^{\o}(t)  \big| \cr
&=& O_p(N^{-\delta/2})+O_p(N_1/N) \cr
&& +O_p( (\log N)^{1/4}/T^{1/2})+O_p(N^{-\nu}) \cr
&& +O_p\Big(N^{-1}T\|\bs \mu\|^2\Big)+O_p\Big(N_1^{1/2}N^{-1}T\|\bs \mu\|\Big).
\end{eqnarray}
\end{theorem}
\noindent{Theorem~\ref{Theorem-1} reveals that the \textit{estimation approach} for the
$\FDR$ in \eqref{2.12} is consistent when $\pi_0=1$ for any
$t>0$. We remark here that $\pi_0=1$ is implied by the sparsity condition (C2).
Otherwise, the average of the signals on the alternative $\|\bs\mu\|/\sqrt{N_1}$ is of the order $O(T^{-1})$, which decreases to zero at a faster rate than that under the local alternative, resulting in lower power.
Unlike the non-sparsity case, for any fixed threshold $t>0$, $\FDP^{\o}(t)= V^{\o}(t)/\{ R^{\o}(t) \vee 1\} \rightarrow 1$ when $\pi_0=1$. This phenomenon is not surprising because the $p$-values under the alternative are negligible compared with that under the true null. From this point of view, the FAT procedure is expected to lose FDR control simply by fixing $t$ under the sparsity setting. In other words, to control the FDR at the nominal level $\alpha$, the data-driven threshold $t_{\alpha}(\hat{\FDR}_{\lambda})$
should go to zero, but at a rate slower than $O_p( 2\Phi(-\sqrt{2 \log N}) )$ asymptotically,  due to the fact that $\max_{i\in \mI_0} |\hat{T}_i| = O_p(\sqrt{2\log N})$.
This further motivates us to investigate the behavior of $\widehat{\FDR}_{\lambda}(t)$ by letting $t\rightarrow 0$ at a rate faster than $O_p( 2\Phi(-\sqrt{2 \log N}))$, a domain under which all of the discoveries are true positives and there are no false positives. However, we may miss some skilled or unskilled funds if the corresponding signals are not strong enough.
To identify and recover all of the true positives, we study the selection consistency property in the following theorem.
\begin{theorem} \label{Theorem-2}
Suppose the assumption~\eqref{2.3} and $\mathrm{(C1)}$-$\mathrm{(C5)}$ hold.
If $\pi_0=1$  and $\min_{i \in \mI_1} |\mu_i|  \ge C_1T^{-\kappa}$, for some constants $C_1> 0$
and $\kappa+\hbar/2<1/2$, then there exists a sequence of threshold values $t_T\rightarrow 0$ such that
$\mathrm{P}(\hat\mI_1^{t_T}=\mI_1)\rightarrow1$, where $\hat{\mI}^{t_T}_1=\big\{i; P_i \le t_T\big\}$.
\end{theorem}
\noindent
%The proof is given in Appendix E of a supplementary file.
From \eqref{2.8} and the minimal signal condition on the $\mu_i$ for $i \in \mI_1$, $\min_{i \in \mI_1} |\hat{T}_i|$ would go to infinity at a rate faster than $(\1^\top \mQ(\mX) \1)^{1/2} T^{-\kappa} \propto T^{1/2-\kappa}$, implying $\max_{i \in \mI_1} P_i = O_p( 2\Phi(-T^{1/2-\kappa}) )$. This, together with the fact that $\min_{i \in \mI_0} P_i= O_p( 2\Phi(-\sqrt{\log N}) )$ and condition (C1), results in $\max_{i \in \mI_1}P_i  < \min_{i \in \mI_0} P_i$ when $N, T$ are sufficiently large. Thus, the estimated true null and non-null sets can be distinguished consistently if the threshold values are selected as $t_T= 2\Phi(-C_2T^{\jmath}) $ for some $\hbar/2<\jmath<1/2-\kappa$ and some positive constant $C_2$.
In summary, the proposed FAT not only provides consistent model selection under some
minimum signal assumptions, but can also control the FDR for any pre-specified nominal level of
$\alpha>0$. This finding is quite important, especially in finite samples; see, for example, \cite{Wasserman:Roeder:2009} for a detailed discussion.

\section{Simulation Studies} \label{Sec-4}

To gauge the finite sample performance of the proposed method, we conduct several simulation studies in this section.
We especially focus on two aspects, namely, the FDP and FDR properties of the FAT procedure in finite sample.

\subsection{Simulation models and competing methods}

We simulate model \eqref{2.1}
in the spirit of a standard capital asset pricing model (CAPM); that is,
\begin{equation*}
Y_{i j}=\mu_i+X_j\beta_i+\varepsilon_{i j}.
\end{equation*}
A proportion $\pi_0$ of the intercepts $\{\mu_i, i=1, \ldots, N\}$ are set to be zero, while
the rest are equal to some finite constant
$\mu$. To make the simulation more realistic, the observed factor $X_j$ and the associated parameters $\beta_i$ are calibrated from our real data discussed in section 5 \citep{Fan:Liao:Mincheva:2013}.
Specifically, to generate the observed factor $X_j$,
we evaluate the mean ($\mu_m$) and variance ($\var_m$) of the monthly market return (in the scale of percentage change) ranging from 11/1995 to 08/2013, which leads to
$\mu_m=0.55$ and $\var_m=4.7^2$. Then, $X_j$ is independently generated from $N(\mu_m, \var_m)$. In addition, to generate the factor loadings,
we fit the CAPM with our real data and evaluate the intercept ($\hat\mu$) and the mean ($\mu_b$) and variance ($\var_b$) of $N=764$ estimated factor loadings, which leads to
$\hat\mu=0.2$, $\mu_b=0.94$ and $\var_b=0.2^2$. Accordingly, $\beta_i$ is independently generated from $N(\mu_b, \var_b)$.
We next consider the error term $\varepsilon_{i j}$.
Motivated by the empirical results for mutual fund selection shown in Section~\ref{Sec-5},
we simulate the random error $\varepsilon_{i j}$ from a latent factor model
with only one common factor; that is,
\begin{equation*}
\varepsilon_{ij}=\gamma_i Z_j+\eta_{ij},
\end{equation*}
where $Z_j $ is independently drawn from a standard normal distribution.
To generate $\gamma_i$, we employ the principal component method to the residuals obtained after fitting the CAPM, and calculate the mean ($\mu_{\gamma}$) and variance ($\var_{\gamma}$) of the $N=764$ estimated latent factor loadings and variance of the residuals after the principal component method ($\var_e$), which leads to $\mu_{\gamma}=0.11$, $\var_{\gamma}=1.44^2$ and $\var_e=2.53^2$. Accordingly, $\gamma_i $ is independently drawn from $N(\mu_{\gamma}, \var_{\gamma})$.
Furthermore, $\bs \eta_{.j}=(\eta_{1j}, \ldots, \eta_{Nj})^\top$ is independently sampled from a
multivariate normal distribution with a mean of $\mathbf{0}$ and a covariance matrix
$\Sigma_{\bs \eta}=(\sigma_{\bs \eta, i_1 i_2})_{N \times N}$ with
$\sigma_{\bs \eta, i_1i_2}=\var_e \times\rho^{|i_1-i_2|}$. One can easily verify that the covariance matrix $\Sigma_{\bs \eta}$ satisfies
the weak dependence assumption \eqref{2.3}. Throughout this section, the autoregressive coefficient $\rho$ and the tuning parameter $\lambda$ used for estimating $\pi_0$ are
simply set to $0.5$ and $0.5$, respectively. All of the simulation results are based on $500$ replications.

To illustrate the superiority of our FAT procedure, we compare it to two existing procedures based on the unadjusted $p$-values:
\begin{description}
\item
{\it Unadjusted procedure:} the method without adjusting for any latent common factors \citep{Barras:Scaillet:Wermers:2010}.
In other words, the unadjusted multiple testing procedure applies the method of \cite{Storey:Taylor:Siegmund:2004} to the unadjusted $p$-values, i.e., $\{ P_i^{\u}, i=1, \ldots, N\}$.
\item
{\it $\mathrm{PCA}$-$\mathrm{PFA}$:} according to the principle factor approximation (PFA) in \cite{Fan:Han:Gu:2012} and \cite{Fan:Han:2013}, the $\FDP$ for the unadjusted $p$-values (i.e., $P_i^{\u}$) with respect to a threshold $t$ can be approximated as
\begin{eqnarray*}
 \widehat{ \FDP}^{\u}(t)& =& \min \Big(
\sum_{i=1}^{N} \Bigg[
 \Phi \Big\{ \frac{  (\hat \sigma_{\mE, ii} )^{1/2} z_{t/2}+ (\1^\top \mQ(\mX) \1)^{-1/2}\1^\top \mQ(\mX) \hat{\mbf Z}\hat{\bs \gamma}_{i} }
 { ( \hat{\sigma}_{\bs \eta, ii})^{1/2}  }  \Big\}  \cr
&&+ \Phi \Big\{\frac{  (\hat \sigma_{\mE, ii})^{1/2}z_{t/2}- (\1^\top \mQ(\mX) \1)^{-1/2} \1^\top \mQ(\mX) \hat{\mbf Z}\hat{\bs \gamma}_{i} }
{  (\hat\sigma_{\bs \eta, ii} )^{1/2}   } \Big\}
 \Bigg] ,R^{\u}(t)  \Big) / R^{\u}(t),
\end{eqnarray*}
where $\hat{\sigma}_{\mE, ii}$, $\hat{\sigma}_{\bs\eta, ii}$, $\hat{\mbf Z}$, and $\hat{\bs \gamma}_i$ are the same as in our FAT procedure, and $R^{\u}(t)=\# \{ i \in \{1, \ldots, N\}: P_i^{\u} \le t \}$. For a pre-chosen $\alpha$, the data-driven threshold is
determined by
\begin{equation*}
t_{\alpha}(\widehat{\FDP}^{\u})=\sup \big\{ 0 \le  t \le 1;  \  \widehat{\FDP}^{\u}(t) \le \alpha \big\}.
\end{equation*}
\end{description}
Note that both the unadjusted method and the PCA-PFA procedure are used to estimate the FDP based on the unadjusted $p$-values ($P_{i}^{\u}$), while our adjusted procedure is derived to estimate the true FDP corresponding to the Oracle factor-adjusted $p$-values ($P^{\o}_i$).

\subsection{FDP comparison: from negative dependence to consistency}
To evaluate how the FDP performs in the $\FAT$ procedure, we consider a scenario with $N=2,000$, $T=215$, and $\pi_0=0.90$, where the number of latent factors is selected via the eigenvalue ratio test in \cite{Wang:2012} and \cite{Ahn:Horenstein:2013} since the method of \cite{Bai:Ng:2002} depends on the choice of tuning parameters. We adopted the criterion as shown in the Figure 2 of \cite{Fan:Han:Gu:2012} by comparing the estimated FDP and the true one using a 45 degree line. Figure~\ref{Figure-3} depicts the scatter plots of the estimated versus the true FDP for the FAT procedure and the competing procedures when $t=0.01$ and $\mu=0.2, 0.3, 0.5$ are calibrated from the real data as shown in Section~\ref{Sec-5}, respectively. The results can be summarized as follows. From panel (d), the estimated $\widehat{\FDR}_{\lambda}(t)$ for the FAT procedure follows the correct pattern of $\FDP^{\o}(t)$ with little variability, which is consistent with the theoretical finding of Theorem~\ref{Theorem-1} that our \textit{estimation approach} \eqref{2.12} is valid. With a large signal (relative to the sample size $T$), our estimated FDP and the corresponding true one [i.e., $\FDP^{\o}(t)$] seem to follow a line with a negative slope. This ``negative" phenomenon occurs when the signals in the alternative are sufficiently strong such that the $p$-values under the alternative are all smaller than or equal to the threshold $t$. Interested readers can refer to ``the negative dependence of the BH-type estimator in finite sample" in a supplementary file for the rationale. However, this does not violate the consistency of our FDP estimator, because our consistency result in Theorem~\ref{Theorem-1} is based on the deviation between the estimated and the true value, a quantity that is analogous to the length of the ``negative" line, and as $N$ increases, the length decreases. As a comparison, the unadjusted procedure [Panels (b), (e), \& (h)] exhibits a long-tailed negatively correlated pattern between the estimated and true values [$\FDP^{\u}(t)$]. PCA-PFA can track the correct pattern of $\FDP$ but with large variability when the noise $(\mE_j)$ is strongly dependent. In summary, the FAT procedure subtracts the dependent variations shared across the tests from the unadjusted test statistics, and thus provides a more stable estimator of $\FDP$ for a wide range of signals.
\begin{figure}[!ht] %[ptbh]
\centering
\begin{tabular}{c}
\vspace{3mm}
\includegraphics[scale=0.8]{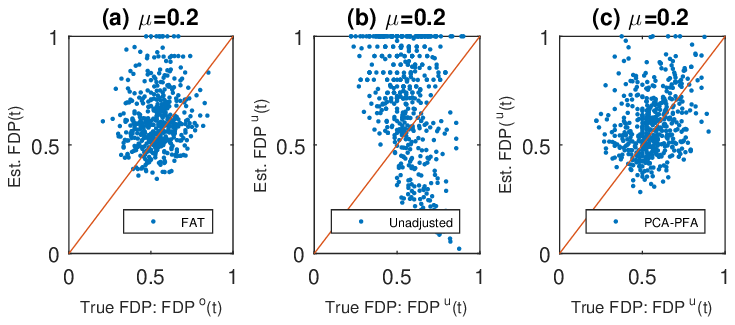}\\
\vspace{3mm}
\includegraphics[scale=0.8]{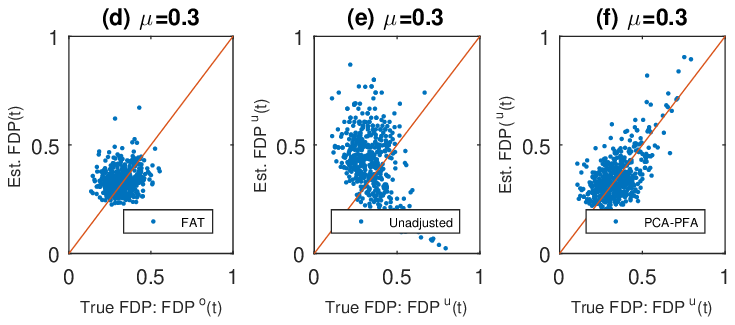}\\
\vspace{3mm}
\includegraphics[scale=0.8]{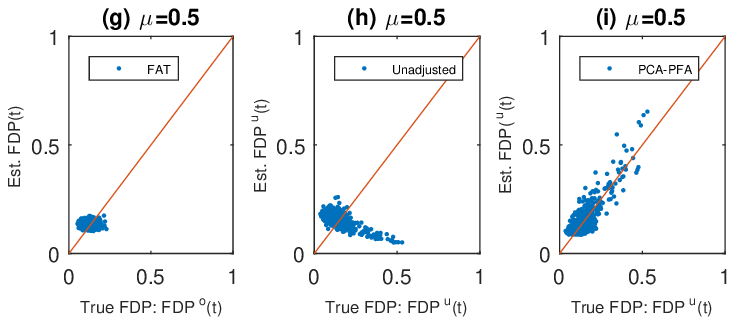}\\
\end{tabular}
\begin{singlespace}
\caption{\textsl{Scatter plots of the estimated $\FDP$ and the true $\FDP$ as a function of $\mu$,  for the $\FAT$, unadjusted, and $\mathrm{PCA}$-$\mathrm{PFA}$ procedures, respectively. The true $\FDP$  for the $\FAT$ is calculated using the oracle factor-adjusted $p$-values, while the true $\FDP$ for the Unadjusted and $\PFA$ methods are calculated using the unadjusted $p$-values. Here, $t=0.01$, $N=2,000$, $\pi_0=0.90$, and $T=215$.} }
\label{Figure-3}
\end{singlespace}
\end{figure}

\subsection{FDR control and power comparison}
To further validate that the data-driven threshold $t_{\alpha}(\widehat{\FDR}_{\lambda} )$ provides strong control of the FDR, Figure~\ref{Figure-4} examines the control of the FDR as well as the power of the FAT, unadjusted, and PCA-PFA procedures for different choices of $\mu$ when $N=2,000$, $T=215$, and $\pi_0=0.90$.
The left panel of Figure~\ref{Figure-4} depicts the empirical $\FDR$
[i.e., the average of 500 $\FDP( \hat{t}_{\alpha})$] for all settings, while the right panel corresponds to the empirical power [i.e., the average of 500 $S(\hat{t}_{\alpha})/N_1$].
Clearly, the empirical $\FDR$s of our $\FAT$ procedure are all controlled at the nominal level $\alpha=0.05$, whereas the empirical $\FDR$s of the unadjusted procedure and PCA-PFA procedure are out of control in some scenarios. Furthermore, the $\FAT$ procedure continues to be more powerful than the competing procedures even when these procedures lose control of the FDR.
The results are expected because subtracting common factors leads to a higher
signal-to-noise ratio.
\begin{figure}[!ht] %[ptbh]
\centering
\includegraphics[scale=0.9]{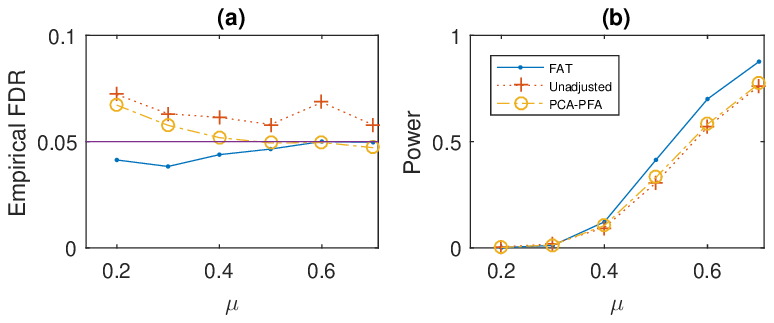}
\begin{singlespace}
\caption{\textsl{Empirical $\FDR$ and power as a function of $\mu$,
for the $\FAT$, unadjusted, and $\mathrm{PCA}$-$\mathrm{PFA}$
procedures, respectively. Here, $\alpha=0.05$, $N=2,000$, $T=215$, and $\pi_0=0.90$. The results are averaged based on $500$ replications. } }
\label{Figure-4}
\end{singlespace}
\end{figure}

\subsection{Model selection consistency}
Under sparsity, it is numerically interesting to show that there exits a sequence of $\alpha_T \rightarrow 0$
such that the data-driven thresholds $t_{\alpha_T}(\widehat{\FDR}_{\lambda} ) \rightarrow 0$ can distinguish the
true null and non-null consistently, in the sense that the power is approaching $1$, while the empirical FDP is shrinking to zero.
To this end, we consider an extremely sparse design with $N=2,000, \pi_0=0.995$, and $\mu=1$. For each $T$, $\alpha_T$ is selected by
\begin{equation}
\alpha_T=\inf \big{\{}  0 \le \alpha \le 1; \  S\{ t_{\alpha}(\widehat{\FDR}_{\lambda} )  \}/N_1 =1  \big{\}}.
\end{equation}
Table~\ref{Table-1} summarizes the average values of $\alpha_T$, $t_{\alpha_T}(\widehat{\FDR}_\lambda ) $, and the empirical $\FDP$ and power for a sequence of $T$ values. As expected,
both $\alpha_T$ and  $t_{\alpha_T}(\widehat{\FDR}_\lambda ) $ are shrinking to zero as
$T$ increases, while the empirical power is always one, from which
Theorem~\ref{Theorem-2} is verified. Moreover, the empirical $\FDP$s are close to $\alpha_T$. These confirm that the FAT procedure can achieve model selection consistency under a sparse setting.
\begin{table}[!ht] %[ptbh]
\begin{center}
\begin{tabular}{c   c c c c } \hline
$T$ & $\alpha_T$ &$\FDP$ &$t_{\alpha_T}(\widehat{\FDR}_\lambda )$   & Power    \\ \hline
200	&0.071938	&0.068865	&0.001835	&1 \\
220	&0.036110	&0.037544	&0.000438	&1 \\
240	&0.019808	&0.019469	&0.000295	&1 \\
260	&0.014148	&0.010459	&0.000082	&1 \\
280	&0.007914	&0.009202	&0.000088	&1 \\
300	&0.003992	&0.002730	&0.000037	&1 \\ \hline
% lambda = 0.5
%200	&0.08288	&0.07810	&0.00124	&1 \\
%220	&0.04439	&0.04270	&0.00069	&1 \\
%240	&0.02239	&0.02130	&0.00015	&1 \\
%260	&0.01138	&0.01167	&0.00007	&1 \\
%280	&0.00816	&0.00805	&0.00007	&1 \\
%300	&0.00464	&0.00504	&0.00004	&1 \\ \hline
\end{tabular}
\caption{\textsl{The performance of $t_{\alpha_T}(\widehat{\FDR}_\lambda ) $ under sparsity.} } \label{Table-1}
\end{center}
\end{table}

\section{Real Data Analysis} \label{Sec-5}
\subsection{Background and data description}

To further illustrate the utility of the proposed FAT procedure,
we consider a real data analysis intended
to pick skilled funds in the U.S. financial market. The data are from the Bloomberg database, one of the most popular and authoritative databases in the world. After
eliminating the funds with missing values, we finally collected
a total of $T=215$ observations on $N=764$ mutual funds in the U.S.
financial market from 11/1995 to 08/2013, with each observation corresponding to one particular fund's monthly excess return $(Y_{i j})$, defined as the fund return $(r_{\mathrm{mj}})$ minus the risk free interest rate $(r_{\mathrm{fj} })$. Here, $r_{\mathrm{fj}}$ is proxied by the monthly 30-day T-bill of the beginning of the month yield. In this dataset, $N=764$ is much larger than $T=215$, which is in accordance with our theoretical findings and
simulation studies. According to the definition of \cite{Barras:Scaillet:Wermers:2010}, a fund can be regarded as a skilled fund if its risk-adjusted return, which is defined as $\mu_i$ in model \eqref{2.4}, is larger than zero. In contrast, a fund is an un-skilled fund if $\mu_i<0$, while the funds with $\mu_i=0$ are called zero-alpha funds; that is, their fund returns are mainly due to noise.
According to the above definition, the main focus is to identify these funds with
$\mu_i>0$ in model \eqref{2.4} involving some observed and unobserved systematic risks, which can be achieved using our FAT procedure.

\subsection{Trading strategy}

We use the $\mathrm{CAPM}$ of \cite{Sharpe:1964} and \cite{Lintner:1965} as the benchmark pricing model; that is,
the observable systematic risk is the single market risk $r_{\mathrm{mj}}-r_{\mathrm{fj}}$, which is proxied by
the excess return on CRSP NYSE/Amex/NASDAQ value-weighted returns for month $j$. It is noteworthy that the reason we use an observable explanatory variable (i.e., the market risk)
is to make a comparison with the notable CAPM, because the market index is found to be an essential factor
that can influence all stock returns. Our FAT procedure can help us to assess whether the single market index
is adequate. As the above procedure relies on some
observable explanatory variables, for the purpose of illustration, we also
consider the factor adjusted
procedure without considering any observed explanatory variables (named EFAT
hereafter), following one anonymous referee's kind suggestion; i.e., no explanatory variables
are considered in (2.1).
To assess the usefulness of the proposed
FAT procedure, we need to verify that the funds that are selected by our FAT procedure can yield higher returns in the future. To this end,
we consider the following rolling window procedure. Specifically,
for some given length of estimation window $L$, for each observation
$\tau\in\{1, \ldots, T-L\}$, we conduct the proposed FAT procedure
using data from periods $\tau$ to $\tau+L-1$ with a pre-specified $\FDR$ level $\alpha$ to select the funds with $\mu_i\not=0$.
Specifically, for any time period $\tau$ to $\tau+L-1$, let $d_{\tau}$ be the estimated number of latent common factors, $\mS^+_{\tau}$ be the collection of selected funds with $\mu_i>0$, and $\mS^-_{\tau}$ be the collection of selected funds with $\mu_i<0$. Let $|\mS^+_{\tau}|$ and $|\mS^-_{\tau}|$ be the number of elements in $\mS^+_{\tau}$ and $\mS^-_{\tau}$, respectively.
We next form a portfolio based on the selected funds $\mS^+_{\tau}$ and $\mS^-_{\tau}$. From each time from $\tau+L$ to $\tau+L+11$ (the next year), we buy the funds within set $\mS^+_{\tau}$ with equal weight $1/|\mS^+_{\tau}|$, and sell the funds within set $\mS^-_{\tau}$ with weight $1/|\mS^-_{\tau}|$ on each fund; then, the return for this strategy from time $\tau+L$ to $\tau+L+11$ is given by $r^{\mathrm{FAT}}_{\tau}=\sum_{i \in\mS^+_{\tau}}\sum_{j=0}^{11} Y_{i, \tau+L+j}/\{12\times |\mS^+_{\tau}|\}
-\sum_{i\in\mS^-_{\tau}}\sum_{j=0}^{11} Y_{i, \tau+L+j}/\{12\times |\mS^-_{\tau}| \}$.
For comparison, we also report the results for the method without controlling for any of the
latent common factors of \cite{Barras:Scaillet:Wermers:2010} (named NFAT hereafter).
Accordingly, we refer to the return for EFAT and NFAT as $r^{\mathrm{EFAT}}_{\tau}$  and $r^{\mathrm{NFAT}}_{\tau}$ from time $\tau+L$ to $\tau+L+11$, respectively.

\subsection{Performance comparison}

We set $L=120$ (ten years) and $\alpha=0.05$ in this analysis, while the results for other selections of
$L$ and $\alpha$ are quite similar. To compare the results based on the above three trading strategies ($\FAT$, $\mathrm{NFAT}$, and $\mathrm{EFAT}$), we report the following performance measures. For any $\tau\in\{1, \ldots, T-L\}$, we compute the average
latent common factors $\mbox{ACF}=\sum_{\tau} d_{\tau}/(T-L)$, the average proportion of selected skilled funds
$\mbox{ASF}=(T-L)^{-1}\sum_{\tau}|\mS^+_{\tau}|/N$, and the average proportion of selected un-skilled funds
$\mbox{ANSF}=(T-L)^{-1}\sum_{\tau}|\mS^-_{\tau}|/N$.
To compare the return persistency of these three trading strategies, we set $\mbox{NFAT}$ as the benchmark and
collect the
return difference of the other two methods with $\mbox{NFAT}$
as $r^{\mathrm{diff-FAT}}_{\tau}=r^{\mathrm{FAT}}_{\tau}-r_{\tau}^{\mathrm{NFAT}}$
and $r^{\mathrm{diff-EFAT}}_{\tau}=r^{\mathrm{EFAT}}_{\tau}-r_{\tau}^{\mathrm{NFAT}}$, and then evaluate the mean of
 $r^{\mathrm{diff-FAT}}_{\tau}$ and $r^{\mathrm{diff-EFAT}}_{\tau}$ together with their associated $p$-values.
The results for the three methods across different performance measures are summarized in Table~\ref{tab:rslt}.
Table~\ref{tab:rslt} shows that
the average number of
latent common factors for FAT is 1.75, which implies that other than market risk, there are indeed some other risk factors that affect fund return.
Nevertheless, the number of latent common factors should vary with time.
The average proportions of selected skilled funds for the three methods are
0.46\%, 0\%, and 0\%, while the average proportions of selected un-skilled funds for the three methods are 3.91\%, 0\%, and 5.64\%, respectively.
The results indicate that only the factor-adjusted procedure ($\FAT$) has power to identify skilled funds, whereas the un-adjusted procedure ($\mbox{NFAT}$) tends to identify more spurious un-skilled funds in finite samples.
Moreover, we find that $\mbox{EFAT}$ has no power to identify any skilled and unskilled funds, which indicates that the observed predictor, i.e., market index, is indeed essential for explaining the co-variation of asset returns.
Similar results can be found for the return difference.
Specifically, the mean of $r^{\mathrm{diff-FAT}}_{\tau}$ is 0.4781\% with a $p$-value of $0.0007$, which is significantly below
5\%, and the mean of $r^{\mathrm{diff-EFAT}}_{\tau}$ is 0.1379\% with a $p$-value of 0.484.
The results further imply that the trading strategies based on the $\FAT$ procedure is more likely to select the funds that can earn persistent returns in the future.
In sum, the FAT procedure is more useful for selecting skilled funds than the un-adjusted procedure. Moreover, some observable predictors, such as the market index used in this analysis, are indeed essential for explaining the co-variation of asset returns, which demonstrates the merits of the notable capital asset pricing model of \cite{Sharpe:1964} and \cite{Lintner:1965}}.

\begin{table}[h]
\bc \caption{\label{tab:rslt} The trading results for the FAT, NFAT, and EFAT methods.} {\small
\begin{tabular}{l | c | c |c }
\hline
&  FAT & EFAT & NFAT \\
\hline
ACF            & 1.75    &  1    & --  \\
ASF            & 0.46\%   &  0\%   & 0\%  \\
ANSF           & 3.91\%   &  0\%   & 5.64\% \\
Average return & 0.3402\% &  0\% & -0.1379\%  \\
p-values of return differences      & 0.0007   &  0.484   & --   \\
\hline
\end{tabular}}
\ec
\end{table}

\section{Concluding Remark}\label{Sec-6}

In this article,  we propose a factor-adjusted multiple testing procedure for deciding whether the intercept of each unit equals zero in a linear factor model involving some observable and unobservable common factors, while the idiosyncratic errors are allowed to be correlated across units.
Based on the principal component method \citep{Wang:2012} for extracting the unknown common factors, we show theoretically that under some mild conditions, the proposed method can estimate the true FDP consistently for any given threshold. Finally, we show that the $\FAT$ procedure enjoys model selection consistency.

To broaden the usefulness of the proposed method, we conclude the article by identifying two possible
research avenues. As we only test the intercepts in linear factor models, it should be practically useful and theoretically demanding to adapt the proposed method to test the significance of certain regression coefficients.
Another avenue is to extend the proposed method to nonlinear factor models, so that
the response variables are allowed to be discrete.

\scsection{Supplementary Material}
Detailed proofs of Proposition~\ref{Proposition-1}, Proposition~\ref{Proposition-2}, Theorem~\ref{Theorem-1} and Theorem~\ref{Theorem-2} can be found in a supplementary file.

\scsection{Acknowledgements}
Wei Lan is from the Statistics School and Center of Statistical Research, Southwestern University of Finance and Economics (E-mail: facelw@gmail.com). Lilun Du is the
corresponding author and an assistant professor of the Department of ISOM, Hong Kong University of Science and Technology (E-mail: dulilun@ust.hk). Wei Lan's research was supported by National Natural Science Foundation of China (NSFC, 11401482, 71532001).
Lilun Du's research was partially supported by IGN15BM04, SBI16BM01, and Hong Kong RGC ECS26301216. The authors are grateful to the Editor, the AE, and two anonymous referees for their insightful comments and constructive suggestions.

%\bibliographystyle{asa}
%\bibliography{D:/WORKING/RESEARCH/LATEX/reference}
\scsection{REFERENCES}
\begin{description}
\newcommand{\enquote}[1]{``#1''}
\expandafter\ifx\csname natexlab\endcsname\relax\def\natexlab#1{#1}\fi

\bibitem[{Ahn and Horenstein(2013)}]{Ahn:Horenstein:2013}
Ahn, S.~C. and Horenstein, A. (2013).
\enquote{Eigenvalue ratio test for the number of factors,}
\textit{Econometrica}, 81, 1203--1227.

\bibitem[{Bai(2003)}]{Bai:2003}
Bai, J. (2003).
\enquote{Inferential theory for factor models of large dimensions,}
\textit{Econometrica}, 71, 135--171.

\bibitem[{Bai and Ng(2002)}]{Bai:Ng:2002}
Bai, J. and Ng, S. (2002).
\enquote{Determining the number of common factors in approximate factor model,}
\textit{Econometrica}, 70, 191--221.

\bibitem[{Barras et al.(2010)}]{Barras:Scaillet:Wermers:2010}
Barras, L., Scaillet, O. and Wermers, R. (2010).
\enquote{False discoveries in mutual fund performance: measuring luck in estimated alphas,}
\textit{Journal of Finance}, 65, 179--216.

\bibitem[{Benjamini and Hochberg(1995)}]{Benjamini:Hochberg:1995}
Benjamini, Y. and Hochberg, Y. (1995).
\enquote{Controlling the false discovery rate: a practical and powerful
  approach to multiple testing,}
\textit{Journal of the Royal Statistical Society, Series B (Statistical Methodology)}, 57, 289--300.

\bibitem[{Benjamini and Yekutieli(2001)}]{Benjamini:Yekutieli:2001}
Benjamini, Y. and Yekutieli, D. (2001).
\enquote{The control of false discovery rate in multiple testing under dependency,}
\textit{Annals of Statistics}, 29, 1165--1188.

\bibitem[{Berk and Green(2004)}]{Berk:Green:2004}
Berk, J.~B. and Green, R.~C. (2004).
\enquote{Mutual fund flows and performance in rational markets,}
\textit{Journal of Political Economy}, 112, 1269--1295.

\bibitem[{Carhart(1997)}]{Carhart:1997}
Carhart, M. (1997).
\enquote{On persistence in mutual fund performance,}
\textit{Journal of Finance}, 52, 57--82.

\bibitem[{Chamberlain and Rothschild(1983)}]{Chamberlain:Rothschild:1983}
Chamberlain, G, and Rothschild, M. (1983), \enquote{Arbitrage, factor structure, and
mean variance analysis on large asset markets,}
\textit{Econometrica}, 51, 1281--1304.

\bibitem[{Cornell(2009)}]{Cornell:2009}
Cornell, B. (2009).
\enquote{Luck, skill, and investment performance,}
\textit{Journal of Portfolio Management}, 35, 131--134.

%\bibitem[{Cuthbertson et al.(2008)}]{Cuthbertson:Nitzsche:Sullivan:2008}
%Cuthbertson, K., Nitzsche, D. and O'Sullivan, N. (2008).
%\enquote{UK mutual fund performance: skill or luck?,}
%\textit{Journal of Empirical Finance,} 15, 613--634.

\bibitem[{Fama and French(1993)}]{Fama:French:1993}
Fama, E.~F. and French, K.~R. (1993).
\enquote{Common risk factors in the return on stocks and bonds,}
\textit{Journal of Financial Economics}, 33, 3--56.

\bibitem[{Fama and French(2010)}]{Fama:French:2010}
Fama, E.~F. and French, K.~R. (2010).
\enquote{Luck versus skill in the cross section of mutual fund returns,}
\textit{Journal of Finance}, 65, 1915--1947.

\bibitem[{Fan and Han(2017)}]{Fan:Han:2013}
Fan, J. and Han, X. (2017).
\enquote{Estimation of false discovery proportion with unknown dependence,}
\textit{Journal of the Royal Statistical Society: Series B (Statistical Methodology)}, to appear.

\bibitem[{Fan et al.(2012)}]{Fan:Han:Gu:2012}
Fan, J., Han, X. and Gu, W. (2012).
\enquote{Estimating false discovery proportion under arbitrary covariance dependence,}
\textit{Journal of the American Statistical Association,} 107, 1019--1035.
\bibitem[{Fan et al.(2013)}]{Fan:Liao:Mincheva:2013}
Fan, J., Liao, Y. and Mincheva, M. (2013).
\enquote{Large covariance estimation by thresholding principal orthogonal complements,}
\textit{Journal of the Royal Statistical Society: Series B (Statistical Methodology)}, 75, 603--680.

\bibitem[{Friguet et al.(2009)}]{Friguet:Kloareg:Causeur:2009}
Friguet, C., Kloaereg, M. and Causeur, D. (2009).
\enquote{A factor model approach to multiple testing under dependence,}
\textit{Journal of the American Statistical Association,} 104, 1406--1415.

\bibitem[{Gagnon-Bartsch and Speed(2012)}]{Gagnon-Bartsch:Speed:2012}
Gagnon-Bartsch, J.~A. and Speed, T.~P. (2012).
\enquote{Using control genes to correct for unwanted variation in microarray data,}
\textsl{Biostatistics}, 13, 539--552.

\bibitem[{Genovese et al.(2006)}]{Genovese:Roeder:Wasserman:2006}
Genovese, C.~R., Roeder, K. and Wasserman, L. (2006).
\enquote{False discovery control with p-value weighting,}
\textit{Biometrika}, 93, 509--524.

%\bibitem[{Grinblatt and Titman(1992)}]{Grinblatt:Titman:1992}
%Grinblatt, M. and Titman, S. (1992).
%\enquote{The persistence of mutual fund performance,}
%\textit{Journal of Finance,} 47, 1977--1984.

%\bibitem[{Hanck(2009)}]{Hanck:2009}
%Hanck, C. (2009).
%\enquote{For which countries did PPP hold? A multiple testing approach,}
%\textit{Empirical Economics}, 37, 93--103.

%\bibitem[{Jin et al.(2014)}]{Jin:Su:Zhang:2014}
%Jin, S., Su, L. and Zhang, Y. (2014).
%\enquote{Nonparametric testing for anomaly effects in empirical asset pricing models,}
%\textit{Empirical Economics}, Forthcoming.

\bibitem[{Kleibergen and Zhan(2013)}]{Kleibergen:Zhan:2013}
Kleibergen, K. and Zhan, Z. (2013).
\enquote{Unexplained factors and their effects on second pass R-square's and t-tests,}
\textit{Working Paper}.

\bibitem[{Leek and Storey(2008)}]{Leek:Storey:2008}
Leek, J.~T. and Storey, J.~D. (2008).
\enquote{A general framework for multiple testing dependence,}
\textit{Proceedings of the National Academy of Sciences of United States of America,} 105, 18718--18723.

\bibitem[{Lintner(1965)}]{Lintner:1965}
Lintner, J. (1965).
\enquote{The valuation of risk assets and the selection of risky investment
in stock portfolios and capital budges,} \textit{Review of Economics and Statistics}, 47, 13--37.

%\bibitem[{Lyons(1988)}]{Lyons:1988}
%Lyons, R. (1988).
%\enquote{Strong laws of large numbers for weakly correlated random variables,}
%\textit{Michigan Mathematical Journal}, 35, 353--359.

%\bibitem[{Pesaran(2006)}]{Pesaran:2006}
%Pesaran, M.~H. (2006).
%\enquote{Estimation and inference in large heterogeneous panels with a
%multi-factor error structure,} \textit{Econometrica}, 74, 967--1012.

\bibitem[{Sharpe(1964)}]{Sharpe:1964}
Sharpe, W.~F. (1964).
\enquote{Capital asset prices: a theory of market equilibrium under conditions of risk,}
\textit{Journal of Finance}, 19, 425--442.

\bibitem[{Storey et al.(2004)}]{Storey:Taylor:Siegmund:2004}
Storey, J.~D., Taylor, J.~E. and  Siegmund, D. (2004).
\enquote{Strong control, conservative point estimation and simultaneous conservative
consistency of false discovery rates: a unified approach,}
\textit{Journal of the Royal Statistical Society, Series B}, 66, 187--205.

\bibitem[{Sun and Cai(2009)}]{Sun:Cai:2009}
Sun, W. and Cai, T. (2009).
\enquote{Large-scale multiple testing under dependence,}
\textit{Journal of the Royal Statistical Society: Series B (Statistical Methodology)}, 71, 393--424.

\bibitem[{Wang(2009)}]{Wang:2009}
Wang, H. (2009).
\enquote{Forward regression for ultra-high dimensional variable screening,}
\textit{Journal of the American Statistical Association}, 104, 1512--1524.

\bibitem[{Wang(2012)}]{Wang:2012}
Wang, H. (2012).
\enquote{Factor profiled independence screening,}
 \textit{Biometrika}, 99, 15--28.

\bibitem[{Wasserman and Roeder (2009)}]{Wasserman:Roeder:2009}
Wasserman, L. and Roeder, K. (2009).
\enquote{High dimensional variable selection,}
\textit{Annals of Statistics}, 37, 2178--2201.

\end{description}

\end{document}